\pdfoutput=1
\documentclass[a4paper,12pt]{article}
\usepackage[utf8]{inputenc}
\usepackage{amssymb,amsmath,amsfonts,bbm,dsfont,hyphenat,bm,float,a4wide,amsthm,amscd,amssymb,slashed}
\usepackage{libertine}
\usepackage{enumerate}
\usepackage[left=2cm,top=3cm,right=2cm,bottom=3cm]{geometry}
\usepackage[british]{babel}
\usepackage{xcolor}
\definecolor{rosso}{cmyk}{0,1,1,0.55}
\usepackage[colorlinks,linkcolor=rosso,citecolor=rosso,urlcolor=rosso]{hyperref}
\usepackage[square,numbers,sort&compress]{natbib}
\usepackage{subfig}
\usepackage{mathtools}
\usepackage{import}
\usepackage[font=small,labelfont=bf,justification=raggedright]{caption}
\def\bee{\begin{equation}}
\def\eee{\end{equation}}
\def\bn{\boldsymbol{n}}
\def\be{\boldsymbol{e}}

\def\zee{h}
\def\anis{u}

\newcommand{\beq}{\begin{eqnarray}}
\newcommand{\eeq}{\end{eqnarray}}

\DeclareMathOperator{\U}{U}

\DeclareMathOperator{\Og}{O}
\DeclareMathOperator{\SO}{SO}

\newcommand{\p}{\partial}
\renewcommand{\i}{\mathrm{i}}
\renewcommand{\d}{\mathop{}\!\mathrm{d}}

\newcommand{\calE}{\mathcal{E}}

\newcommand{\bN}{\bm{N}}

\newcommand{\nub}{\bar{\nu}}
\newcommand{\pb}{\bar{\partial}}

\newtheorem{theorem}{Theorem}

\newtheorem{definition}[theorem]{Definition}

\usepackage{tikz}
\usetikzlibrary{decorations.pathreplacing}

\title{\textcolor{rosso}{Moduli spaces and breather dynamics of analytic solutions in Heisenberg exchange-free chiral magnets}}

\author{Bruno Barton-Singer,$^{1,}$\thanks{\textcolor{rosso}{bbarton$\_$singer@iacm.forth.gr}} \ \ Stefano Bolognesi,$^{2,}$\thanks{\textcolor{rosso}{stefano.bolognesi@unipi.it}} \\  Sven Bjarke Gudnason,$^{3,}$\thanks{\textcolor{rosso}{Corresponding author: gudnason@henu.edu.cn}} \ \ Roberto Menta$^{4,}$\thanks{\textcolor{rosso}{roberto.menta@sns.it}}
\\[13pt]
  {\footnotesize
    $^{1}$Institute of Applied and Computational Mathematics,
Foundation for Research}\\[-5pt]
   {\footnotesize and Technology - Hellas, 700 13 Heraklion, Greece}\\[2pt]
  {\footnotesize
    $^{2}$Department of Physics ``E. Fermi'', University of Pisa, and INFN, Sezione di Pisa,}\\[-5pt]
  {\footnotesize
    Largo Pontecorvo, 3, Ed. C, 56127 Pisa, Italy}\\[2pt]
  {\footnotesize	$^{3}$Institute of Contemporary Mathematics, School of Mathematics and Statistics,
  }\\[-5pt]
  {\footnotesize
    Henan University, Kaifeng, Henan 475004, P.~R.~China
  }\\[2pt]
  {\footnotesize
    $^{4}$Scuola Normale Superiore, Piazza dei Cavalieri, 7, and Laboratorio NEST, 
    }\\[-5pt]
  {\footnotesize
    Piazza S.~Silvestro, 12, 56127  Pisa, Italy}\\[2pt]
}
\date{\small\today}

\begin{document}

\maketitle

\abstract{
We investigate the special case of the chiral magnet with vanishing
Heisenberg exchange energy, whose axisymmetric Skyrmion solution has
previously been found. The dynamical equations of this model look like
inviscid fluid flow, and by investigating path lines of this flow we
can construct explicit static and dynamic solutions. We find an
infinite-dimensional family of static Skyrmions that are related to
the axisymmetric Skyrmion by co-ordinate transformations thus
discovering a new moduli space, and further infinite-dimensional
families of axisymmetric and non-axisymmetric breather-like
supercompactons.
}

\newpage
\section{Introduction}

The magnetic Skyrmion is a spin texture in two spatial dimensions stabilized by the chiral interaction known as Dzyaloshinskii-Moriya interaction (DMI) \cite{dzyaloshinsky1958thermodynamic,moriya1960anisotropic} as well as an external potential in the form of the Zeeman energy or the anisotropy term \cite{bogdanov1989thermodynamically}.
The DMI is related to the spin-orbit interaction and breaks parity, 
energetically favouring either Skyrmions or anti-Skyrmions, but not both.
The DMI term is present in chiral magnets and thin films thereof and indeed the magnetic Skyrmion has first been experimentally discovered in MnSi \cite{doi:10.1126/science.1166767}.
The DMI tends to enlarge the solitonic texture -- the Skyrmion \cite{nagaosa2013topological}, whereas the potential tends to shrink it. The balance of these two forces gives the size of the Skyrmion.
The Heisenberg exchange energy (relativistic kinetic energy in $\sigma$-model language) on the other hand plays no role in this balance of forces -- it is classically conformally invariant in two dimensions. 
Nevertheless, the Heisenberg energy ensures that the soliton is completely smooth, at least in the continuum limit.

The single magnetic Skyrmion possesses radial symmetry in the sense that its energy is axisymmetric. The field is axisymmetric in the sense that as a vector field it is invariant under rotation. The phase of the Skyrmion is fixed by the chiral interaction to minimize the energy of the DMI, such that its integral is negative.
Although the texture is axisymmetric and the profile is described by a simple second-order differential equation, its solution cannot be written explicitly in closed form.
It is quite rare that such equations have analytic solutions, with the exception of integrable or sometimes supersymmetric models.
For superconductors, the latter are known as critically coupled vortices, but even these do not possess a known analytic solution.

In the case of magnetic Skyrmions, several exact analytic solutions are known in certain limits of the general theory.
The BPS (Bogomol'nyi-Prasad-Sommerfield) or supersymmetric limit possesses many solutions, including the single magnetic Skyrmion \cite{Barton-Singer:2018dlh}, but a special potential of the form of the Zeeman energy squared is required: the coupling of the Zeeman energy (external magnetic field) must match exactly half of the effective anisotropy coupling, and it should also be half of the square of the DMI coupling.
This precise limit is related to BPS solutions in a related gauged non-linear sigma model~\cite{schroers2019gauged, schroers2021solvable}.
The single magnetic Skyrmion solution was found away from this limit by D\"oring and Melcher, where the relation between the the coupling of this potential and DMI coupling is relaxed \cite{Doring2017}.
Finally, in a recent paper by some of the authors \cite{Bolognesi:2024mjs}, several analytic solutions can be applied to the general theory of magnetic Skyrmions in two limits: the lump limit, where the solution is described mainly by the Heisenberg energy and the opposite limit, where the Heisenberg energy is negligible -- the restricted limit.
The exact solution by D\"oring and Melcher is in fact of the lump type, whereas the restricted analytic solutions depend strongly on the potential in question.
The restricted limit of the magnetic Skyrme theory contains only the DMI and a potential and the solutions can be compactons, supercompactons or normal solitons with a tail.
Compactons are solitons with compact support, that usually have discontinuous derivatives of their fields at the compacton boundary -- but a continuous energy density.
Supercompactons on the other hand, are also discontinuous in the fields themselves \cite{Bolognesi:2024mjs}.

One may think that the limit of the theory without the Heisenberg exchange interaction is rather strange.
Nevertheless, it quite naturally arises in certain supersymmetric models, where 
supersymmetrization of the theory ``eats'' the kinetic term \cite{Freyhult:2003zb,Adam:2013awa,Nitta:2014pwa}.
More precisely, this happens for $\mathcal{N}=2$ supersymmetric extensions, because the solution of the auxiliary field contains the kinetic term with the opposite sign, when substituted back into the Lagrangian\footnote{This also happens in $3+1$ dimensions for a Skyrme-like theory \cite{Gudnason:2015ryh}.}. Moreover, exact solutions in a certain limit can be used to approximate solutions in the vicinity of that limit, as for example the solutions in the BPS limit were used to find novel solutions for a range of coupling parameters \cite{kuchkin2020magnetic}.

A property that often comes with BPS or supersymmetric solutions, is the existence of a moduli space -- a space of solutions with the \emph{same} energy.
This is often a result of the large amount of symmetry present in such theories, severely limiting the types of viable potentials.
The moduli space is not just an abstract mathematical notion, but can often be used to approximate dynamics of solitons in the near-BPS limit, by describing them as points on the moduli space instead of as solutions of full field equations.
The reason why this is not trivial at all, is the fact that the moduli space is often endowed with a curved and sometimes complicated metric.
The first example of this was put forward by Manton in the seminal paper \cite{Manton:1981mp} (see also Ref.~\cite{Atiyah:1992if}), where the scattering of BPS monopoles is described by geodesic motion on the 2-monopole moduli space.

In Ref.~\cite{Bolognesi:2024mjs}, axisymmetric analytic solutions were found for the restricted magnetic Skyrmion.
Due to the limitation on the rotation of the magnetic Skyrmion coming from the DMI, it was conceivable that there would not be a nontrivial moduli space.
In this paper, however, we find an infinite-dimensional moduli space. We study the equations of the restricted magnetic Skyrmion from the point of view of fluid paths and find that they are described by perfect circles.
It turns out that the circles need not be concentric, since the centre can ``move'' as the radius changes. The only constraint is that the circles may not intersect each other. 
We thus propose that the restricted magnetic Skyrmion has a moduli space described by the configurations of a ``string'' which has zero tension up to a certain maximum extension. Reintroduction of an infinitesimal Heisenberg exchange energy gives the string a tension.

This paper is organized as follows.
In Sec.~\ref{sec:fluid} we introduce the restricted model and study the moduli space of the analytic solutions from the point of view of fluid paths.
In Sec.~\ref{sec:dynamics}, we consider the time-dependent counterparts that resemble breather-like Skyrmions.
In Sec.~\ref{sec:3d}, we contemplate the extension of the model to three dimensions, but conclude that Hopfions do not exist.
We conclude in Sec.~\ref{sec:conclusion} with a discussion and outlook.
We relegate some supplemental material to an appendix.

\section{Fluid flow dynamics and moduli space}\label{sec:fluid}

Magnetic Skyrmions are described by families of topological maps $\bn:\mathbb{R}^2 \to S^2$, 3-dimensional magnetization vectors $\bn = (n_1,n_2,n_3)$, such that $|\bn|^2=1$ with $n_3$ being the out-of-plane magnetization. These vortex-like configurations in chiral magnetic materials are usually described by a standard energy functional comprising a free energy Heisenberg term ($E_2$), the characteristic DMI term \cite{dzyaloshinsky1958thermodynamic, moriya1960anisotropic} ($E_1$) and a generic external potential ($E_0$) such as the Zeeman interaction term. It is well-known that this model admits nontrivial topological solutions \cite{bogdanov1989thermodynamically,bogdanov1994thermodynamically}.

In this paper, let us consider such standard two-dimensional chiral magnet energy, but in the limit where the Heisenberg exchange energy $E_2 = \frac{1}{2} (\nabla \bn)^2$ goes to zero -- the so-called restricted limit in Ref.~\cite{Bolognesi:2024mjs}, where $\nabla=(\p_1,\p_2,0)$. 
We hence arrive at the energy 
\bee
\begin{array}{rlc}
E(\bn) &= \displaystyle \int  \d^2x \Big(k\bn\cdot(\nabla\times\bn) + V(\bn)\Big) \\
&\equiv E_1 + E_0 \ ,
\label{eq:energy}
\end{array}
\eee
where $k$ is the DMI strength and $V(\bm{n})$ is a generic external potential. 
We can impose the boundary condition that the field must approach a minimum $\bm{N}$ of the potential at ``large'' distances: $\lim_{|\bm{x}|\to\infty}\bm{n}=\bm{N}$.
This effectively point-compactifies the plane to a 2-sphere, giving the topology to the Skyrmions, $\bn:S^2\to S^2$, which are characterized by the second homotopy group $\pi_2(S^2)=\mathbb{Z}$. The element of this group a field $\bn$ represents can be computed from the integral
\beq
Q = \frac{1}{4\pi}\int\d^2x\;\bn\cdot\p_1\bn\times\p_2\bn\ ,
\label{eq:Q}
\eeq
called the `topological charge'. However more broadly, we can suppose that if there is a degenerate circle of minima, the field approaches different minima in different directions at infinity, mapping the `circle at infinity' to the circle of minima with a prescribed winding number. This is typically forbidden by the resulting divergence of the Heisenberg energy term, but here it is allowed. The resulting topology is then described by relative homotopy of the map $(D^2,\partial D^2)\to(S^2,S^1\xhookrightarrow{} S^2)$, and the integral above is no longer integer, but still invariant under continuous transformations respecting the boundary conditions.
Indeed we will shortly see that there are not only compact solutions, but supercompactons, where the topological charge is fractional (see e.g.~Ref.~\cite{Bolognesi:2024mjs}).

The preserved rotational symmetry group of the potential energy term $E_0$ is the internal $\SO(2)_{\rm int}$, corresponding to rotations about the vacuum direction $\bm{N}$, and the spatial rotational symmetry in the $(x, y)$-plane, $\SO(2)_{\rm space}$.  When the DMI term is included, i.e., for the total energy $E = E_1 + E_0$, the independent internal and spatial rotational symmetries are broken. However, a diagonal combination of these symmetries, $\SO(2)_{\rm diag}$, is preserved. This symmetry corresponds to the transformation of $\bn$ as a vector field under co-ordinate rotation. The separate reflections of internal and spatial reflection symmetries are broken to a single $\mathbb{Z}_2$ symmetry, but this one corresponds to the reflection of $(n_1,n_2)$ and $(x,y)$ on orthogonal axes. Taking into account these symmetries of the model, the unbroken symmetry group is at least
\bee
G = \Og(2)_{\rm diag} \ltimes T_2\ ,
\eee
where $T_2$ represents two-dimensional translations, $\Og(2)_{\rm diag}$ includes both $\SO(2)_{\rm diag}$ and the diagonal parity symmetry described above, and $\ltimes$ is the semi-direct product. When the energy involves the Heisenberg energy term $E_2$, this is the whole symmetry group.

Let us now consider undamped Landau-Lifshitz dynamics \cite{Davidson_2001} with respect to this energy:
\bee
\partial_t\bn=\bn\times \frac{\delta E}{\delta \bn} = -2k(\bn\cdot\nabla) \bn+ \bn\times \frac{\d V}{\d \bn} \ .
\label{fluid_flow_eq}
\eee
We can interpret this dynamical equation as a fluid flow equation with material derivative $D_t:=\partial_t + 2k\bn\cdot\nabla$ such that Eq.~\eqref{fluid_flow_eq} reduces to
\bee
D_t \bm{n} = \bn\times \frac{\d V}{\d \bn} \ .
\eee
That means we can switch to a ``Lagrangian'' description of the Landau-Lifshitz dynamics where we imagine spins being advected by a velocity field $\vec{\nu}=(n_1,n_2)$ at speed $2k$, and as we follow a spin, it precesses around the effective magnetic field $\frac{\d V}{\d\bn}$. Indeed we have that the ``convective derivative'' is equal to $2k \vec{\nu} \cdot \nabla$, with $\partial_{3} := 0$.
Explicitly, we can define a path in (2+1)-dimensional spacetime, $\vec{X}(t)$, such that  $\dot{\vec{X}}=2k\vec{\nu}$ and Eq.~\eqref{fluid_flow_eq} reduces to
\bee
\frac{\d}{\d t}\bn(\vec{X}(t),t)=\partial_t\bn + \dot{\vec{X}}\cdot\nabla\bn =D_t\bn=\bn(\vec{X}(t),t)\times \frac{\d V}{\d\bn}\bigg\rvert_{\bn(\vec{X}(t),t)}.
\eee

In the fluid description, $\vec{X}(t)$ is the trajectory of an infinitesimal fluid packet, a ``path line'' in the language of fluid mechanics \cite{Batchelor_2000}. In various cases, we can solve the above equation and explicitly find this path line and the variation of $\bn$ along it. By putting together path lines that fill every point in space and time, we can construct dynamical solutions. This is an example of the general method of characteristics \cite{Courant_Hilbert_1962}, which is generally useful in solving first-order nonlinear equations.

For the rest of this paper, we only consider potentials that are symmetric around $\be_3$ (``axisymmetry''), meaning $V(\bn)=U(n_3)$. In the standard cases, the potential corresponds to the combination of Zeeman interaction with strength $\zee\geq0$ (without loss of generality) and uniaxial anisotropy with strength $\anis$, i.e.
\bee
U(n_3) = \zee(1-n_3) + \anis(1-n_3^2) \ .
\label{eq:standard_pot}
\eee

If we define the complex velocity field $\nu:=n_1+\i n_2$, then we see that the evolution of a spin along the path line $\vec{X}(t)$ is given by:
\begin{align}
&D_tn_3=0 \ , \\ 
&D_t \nu 
= -\i\nu U'(n_3) \ , 
\end{align}
so $n_3(\vec{X}(t),t)$ is constant along a path line and $\nu(\vec{X}(t),t) = -\sqrt{1-n_3^2}e^{-\i U'(n_3)t}$. We can thus explicitly calculate the curve $\vec{X}(t)$ introducing $X=X_1+\i X_2$:
\bee
\dot{X}(t)=-2k\sqrt{1-n_3^2}e^{-\i U'(n_3)t}\implies X(t)=X_c+\underbrace{\frac{-\i2k\sqrt{1-n_3^2}}{U'(n_3)}}_{\i R(n_3)}e^{-\i U'(n_3)t} \ ,
\label{eq:X_pathline_soln}
\eee
i.e.,
\beq
X(t) = X_{c} + R(n_3)\big[\sin(U'(n_3)t) +\i\cos(U'(n_3)t)\big]\ ,
\eeq
so we see that the curve (path line) is generically a circle, with radius $R$ depending on $n_3$. There are two special cases: if $n_3=\pm1$, then the radius is zero and $\vec{X}(t)$ is a point, $X_c$. Notice that in the case of standard potential \eqref{eq:standard_pot}, $U(-1) > U(1) = 0$ so the lowest energy state is $n_3=1$ everywhere. If on the other hand there are stationary points of $U(n_3)$ for $-1<n_3<1$, then the radius is infinite and $\vec{X}(t)$ is a straight line.

When using the method of characteristics to construct solutions to an equation of motion, discontinuities may occur, and in turn solutions can only satisfy the equation of motion at the integral level, leading to shock conditions. In this paper, we focus on constructing static and dynamical solutions that are continuous in time and space, either for all time or up to a singularity. This means in turn that we must generically restrict the domain to a disc, since the `supercompacton' solutions that are found in this regime are not continuous in the whole plane \cite{Bolognesi:2024mjs}.

\subsection{Stationary flows}
We can also look for the special cases when the flow is stationary, 
meaning that at each point in space, the direction of fluid flow remains the same over time. Returning to the conventional way of looking at magnetization dynamics, this means that $n_1$ and $n_2$ are static, and thus the whole magnetization configuration is static.

\begin{figure}[ht]
\begin{center}
\begin{tikzpicture}[scale=1.0]
\draw (-4,0) circle (3);
\draw[->,rosso] (-4,0) --node[anchor=south] {\small $R(n_3)$}(-1,0) ;
\draw[->,thick]        (-4,-3)   -- (-3,-3);
\draw[->,thick]        (-4,3) node[anchor=south] {\small $n_3$ \small{constant}}   -- (-5,3);
\draw [decorate,decoration={brace,amplitude=4pt,mirror,raise=0.3em}]
 (-4,-3)   -- (-3,-3) node[midway,yshift=-1.5em]{\small $\sqrt{1-n_3^2}$};
\draw[->,thick]        (-1,-0)   -- (-1,1);
\draw[->,thick]        (-7,0)   -- (-7,-1);
\draw (4,0) circle (3);
\draw[->,rosso] (4,0) --node[anchor=south] {\small $-R(n_3)$} (7,0) ;
\draw[->,thick]        (4,-3)   -- (3,-3);
\draw[->,thick]        (4,3) node[anchor=south] {\small $n_3$ \small{constant}}   -- (5,3);
\draw [decorate,decoration={brace,amplitude=4pt,mirror,raise=0.3em}]
 (3,-3)   -- (4,-3) node[midway,yshift=-1.5em]{\small $\sqrt{1-n_3^2}$};
\draw[->,thick]        (1,-0)   -- (1,1);
\draw[->,thick]        (7,0)   -- (7,-1);
\end{tikzpicture}
\caption{Streamlines
of constant $n_3$ for positive and negative $R(n_3)$.}
\label{fig:illustration}
\end{center}
\end{figure}
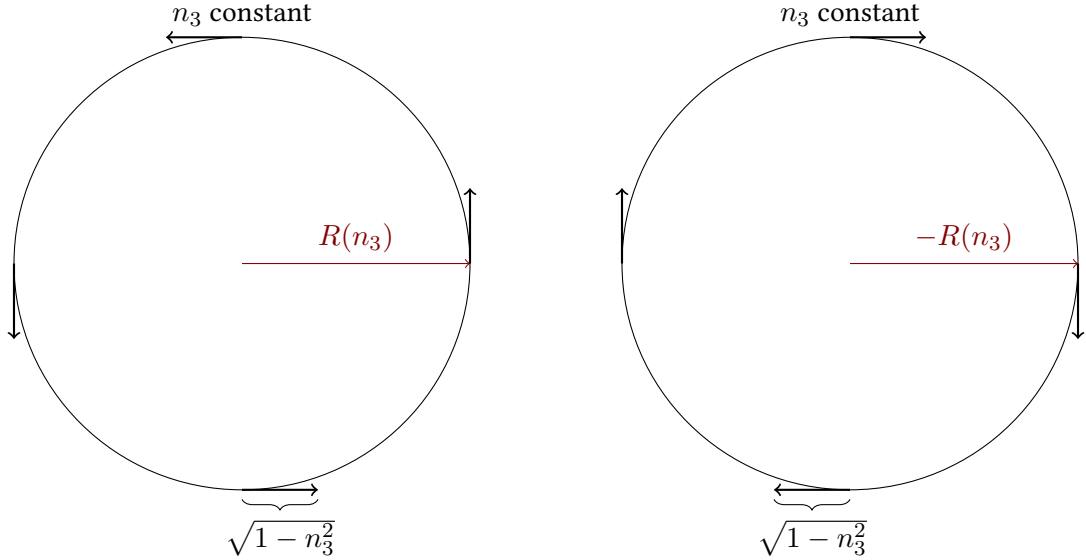

What we have done at this point is to construct a generic path line for fluid flow. If we first specialize to look for stationary solutions, this path line is a streamline, defined as an integral curve of the flow at a given instant in time \cite{Batchelor_2000}. All points on this circle instantaneously have the same $n_3$, and $\nu$ is such that it is always tangent to the circle. By looking at $\lvert \vec{X}(t)-\vec{X}_c\rvert $, the circle will have a signed radius that depends on $n_3$, i.e.
\bee
R(n_3)=\frac{-2k\sqrt{1-n_3^2}}{U'(n_3)} \ .
\label{eq:radius_generic}
\eee
If $R>0$, it indicates a curve with spins tangent to the curve in an anticlockwise direction, conversely $R<0$ in a clockwise direction, as shown in Fig.~\ref{fig:illustration}.

Any static continuous solution to the equation of motion must be assembled from these streamlines in such a way that they do not intersect, and cover the domain. In the case where $R(n_3)$ has a finite maximum $R_{\rm max}$, this domain must be the disc of radius $R_{\rm max}$, or smaller.

\subsection{Static axisymmetric Skyrmion-like solution\label{sec:static_axisymmetric_solns}}

Let us take this a step further and create an axisymmetric 
Skyrmion-like solution to the static Landau-Lifshitz equation \eqref{fluid_flow_eq} by assembling a series of concentric circles (here we set $X_c=0$). For this to be possible, we need $R(n_3)$ to be a monotonic function (or the circles will intersect). We will see below that in general $R(n_3)$ is not monotonic all the way from $n_3=-1$ to $n_3=1$, meaning we can only construct configurations that do not cover the whole target sphere: they have fractional topological charge according to Eq.~\eqref{eq:Q}.

Considering the standard potential \eqref{eq:standard_pot} in Eq.~\eqref{eq:radius_generic}, we get the dependence of radius on $n_3$:
\bee
R(n_3)=\frac{2k\sqrt{1-n_3^2}}{\zee+2\anis n_3} \ .
\label{R_n3}
\eee

\begin{figure}[t]
    \centering
    \includegraphics[width=0.85\linewidth]{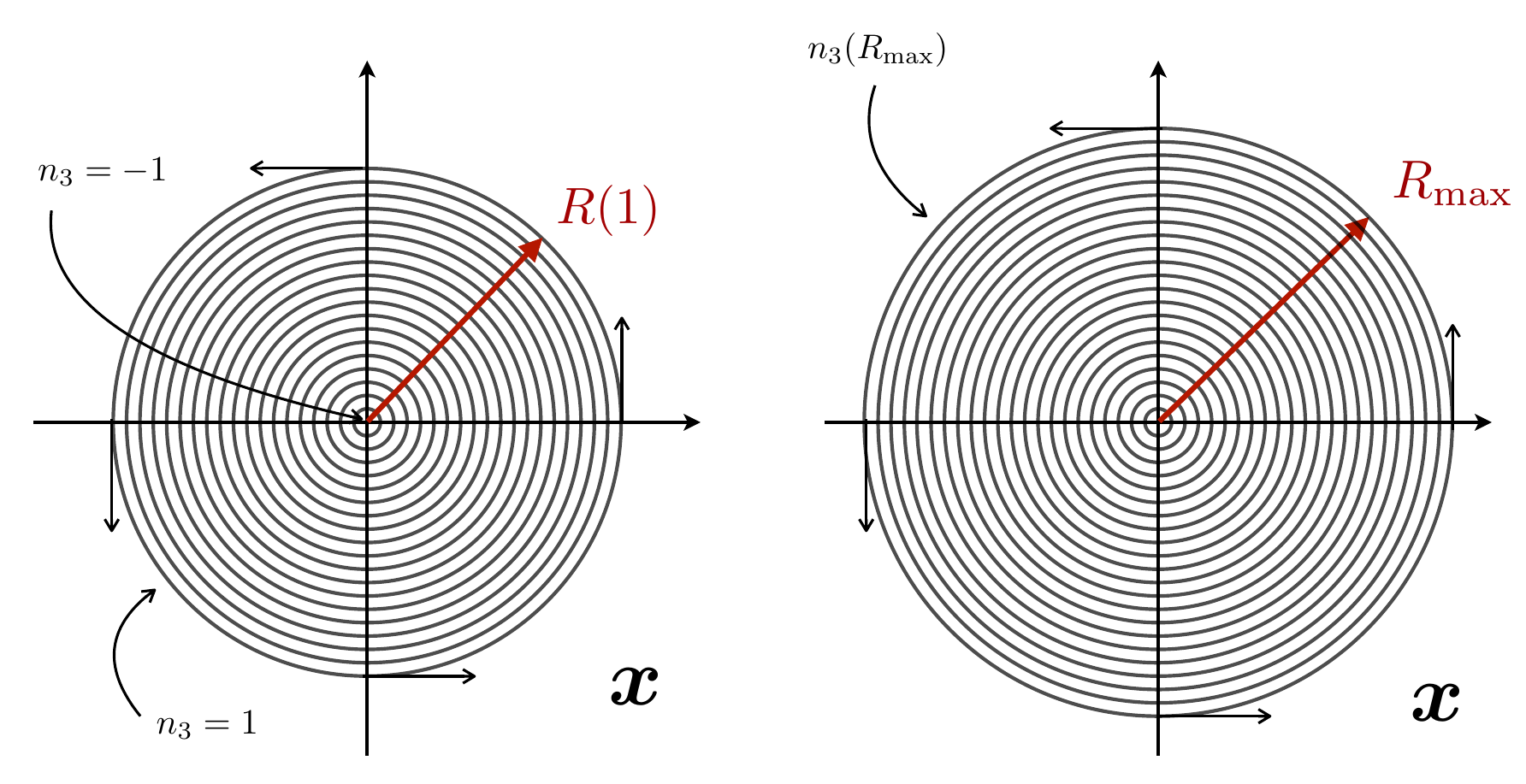}
    \caption{Axisymmetric static Skyrmion-like solution by assembling
      a series of concentric circles for positive $R(n_3)$.
}
    \label{fig:axisymm-skyrm}
\end{figure}

Except for the special cases $\zee=\lvert 2\anis\rvert$, this function is zero at both $-1$ and $1$ and thus cannot be monotonic. However, we see below that when $\zee\neq \lvert 2\anis\rvert$, $R(n_3)$ is monotonic on two intervals $[-1,n_3^\star]$ and $[n_3^\star,1]$, with $n_3^\star$ the location of either a pole or finite maximum of $R$. There are therefore two axisymmetric solutions for a given $(\zee, \anis)$, one with $n_3=-1$ at the centre and the other with $n_3=1$ at the centre, both approaching the same boundary conditions. In case where $n_3^\star$ is the location of a finite maximum, our constructed solution is a `supercompacton' only reaching a finite radius $R_{\rm max}$, so we limit the domain to be a disc of radius $R_{\rm max}$ to avoid discontinuous solutions, as shown in Fig.~\ref{fig:axisymm-skyrm}.

In general, inverting the function \eqref{R_n3} gives the profile
\bee 
n_3(r) = \frac{\rho ^2\tilde{u}-\sqrt{\left(\rho ^2+4\right) \tilde{u}^4+8 \tilde{u}^2-\left(\rho ^2-4\right)}}{\left(\rho ^2+4\right) \tilde{u}^2+4}\ , \qquad \rho=\frac{r}{k}\sqrt{\zee^2 + 4 \anis^2}\ ,\qquad\tilde{u}=\frac{2 u}{h}\ .
\label{solrestricted}
\eee

In the special case $\zee+2\anis =0$, $U'(1)=0$ and therefore $R(1)$ is not necessarily zero. Expanding $n_3 = 1 -\frac{\epsilon^2}{2}$ and substituting in $U'(n_3)$, we find that $ R\to 2k\epsilon /(\anis \epsilon^2)$, so the radius is in fact infinite for $\epsilon\to 0$. Let us explicitly calculate this case:
$R(n_3)=\frac{2k\sqrt{1-n_3^2}}{\zee(1-n_3)}=\frac{2k}{\zee}\sqrt{\frac{1+n_3}{1-n_3}}$, with $n_3=\cos\Theta(r)$ in spherical co-ordinates. We find that
\bee
\Theta(r)=2\arctan\left(\frac{2k}{\zee r}\right),
\eee
is the Bogomol'nyi 
solution of Ref.~\cite{melcher2014chiral}. This makes sense as these solutions had the property that they separately satisfied the Dirichlet and DMI+potential parts of the Euler-Lagrange equation \cite{melcher2014chiral, Bolognesi:2024mjs}, rather than being the solution of a single combined Bogomol'nyi equation as at the critical coupling point~\cite{Barton-Singer:2018dlh}.

Generally, we should first look for the poles of $R$. The condition for a pole is given by $U'(n_3) = 0$, i.e.
\bee
\zee + 2\anis n_3 = 0 \ .
\eee
This implies that in the subset of the polarized phase $\zee-2\lvert \anis\rvert>0$, there are no poles within the range $-1 < n_3 < 1$, meaning we expect finite
$R_{\rm max}$ to be at $n_3^\star<1$, and thus we find supercompactons with fractional topological charge. By contrast, in the canted phase $h+2u<0$ (where $\bN\neq\be_3$) and in the remaining part of the polarized phase $h-2u<0$ we see $R\to \infty$ at $n_3=-\frac{\zee}{2\anis}$. We then have Skyrmions with fractional topological charge that are not supercompactons in both of these regions of the phase diagram, i.e.~they fill the plane and have no discontinuity. To go further, we should look at the maximum of $R(n_3)$:
\bee
R'(n_3)=2k \frac{-( \zee+2\anis n_3) \frac{n_3}{\sqrt{1-n_3^2}}- \sqrt{1-n_3^2} 2 \anis}{(\zee+2\anis n_3)^2}=0\quad\implies\quad n_3^\star=-\frac{2\anis}{\zee}\ ,
\eee
therefore 
\bee
R(n_3^\star)=\frac{2k}{\sqrt{\zee^2-4\anis^2}} \ .
\label{eq:Rn3star}
\eee
The second derivative of $R(n_3)$ evaluated on $n_3^\star$ is equal to 
\bee
R''(n_3^\star) = - \frac{2k}{\zee\left(1 - \frac{4\anis^2}{\zee^2}\right)^{5/2}}\ ,
\eee
which is negative due to $R(n_3^\star)>0$, confirming that $R_{\text{max}} = R(n_3^\star)$.

We can summarize the two regimes:
\begin{equation}
   n_3^\star= \begin{cases}
        \frac{-\zee}{2\anis} & \zee<\lvert 2\anis\rvert\, \text{two solutions in whole plane}\\
        \frac{-2\anis}{\zee}  & \zee>\lvert 2\anis\rvert \, \text{two solutions on disc}
    \end{cases}
\end{equation}
For simplicity, we can refer to these two regimes as `infinite $R_{\rm max}$' and `finite $R_{\rm max}$' respectively.

For the infinite $R_{\rm max}$ solutions, we can ask what the boundary condition at infinity is. When $0 \leq \zee \leq -2u $, $n_3^\star$ is a minimum of $U(n_3)$, meaning these solutions approach the degenerate circle of minima as discussed in Sec.~\ref{sec:fluid}. On the other hand if $0\leq \zee\leq 2\anis$, the field $n_3$ tends to the global maximum of the potential at infinity -- this calls for discarding the solutions in the parameter region $0\leq \zee\leq2\anis$.

As discussed above, the correct topological description of these solutions is relative homotopy. Here we see that at each given $\zee,\anis$ there are two valid static solutions, having the same $+1$ winding around the boundary circle, but one covering the north pole and the other the south pole. Using the formula \eqref{eq:Q}, the topological charge of the configuration with $n_3=\pm1$ at its centre is
\begin{eqnarray}
    Q^{\pm}= \frac{1}{2}\int_0^\infty \sin\Theta\partial_r\Theta \d r = \frac{1}{2}\left(n_3(0)-n_3^\star\right)=\begin{cases}
    \frac{\zee}{\anis}\pm \frac{1}{2} & \zee<\lvert 2\anis\rvert\\
        \frac{\anis}{\zee}\pm\frac{1}{2}& \zee>\lvert 2\anis\rvert
    \end{cases}.\label{eq:top_charge}
\end{eqnarray}
We see that for our solutions, the two distinct relative homotopies always have different topological charges, although in principle topological charge is a cruder measure.

We expect the energy of the two configurations to be different also, but the DMI term is difficult to calculate explicitly.

These radially symmetric solutions can also be reached by imposing radial symmetry and solving the associated first-order ODE.

From the fluids point of view adopted in this paper, the discontinuous solution in the whole plane found in \cite{Bolognesi:2024mjs} can be assembled by putting together this static solution on the disc of radius $R_{\rm max}$ together with a uniform solution on the plane minus the disc, noting that since $\vec{\nu}$ is parallel to the boundary of the disc, there is no shock condition.

\subsection{Infinite-dimensional family of non-axisymmetric solutions}

In the section above, we impose an unnecessary constraint by asking for axisymmetry, which equivalently means that the circular path lines are concentric. We can imagine that by varying the centres of the path lines as a function of the radius, we can construct a large family of solutions with the same energy and topological charge as the axisymmetric one. In this section we ask what is the most general continuous solution we can construct from closed path lines, in general on subsets of the plane bounded by the largest possible path line, if it exists. Such solutions can be put together to make a subset of all discontinuous solutions in the plane.

Trivially, we can construct the uniform state $n_3=1$ or $n_3=-1$ on any domain, since each path line is closed. If we want to construct a solution which differs from the uniform state, it must have path lines that are circles. If the path lines can have arbitrarily large radius, we can again construct a solution in the whole plane. If instead there is a largest possible radius $R_{\text{max}}$, the largest region on which we can define such a solution is the disc $D$ of radius $R_{\text{max}}$, $B(R_{\text{max}})$, with boundary conditions that $\vec{\nu}(\p B(R_{\text{max}}))$ is parallel to the boundary. The axisymmetric solution above is one non-uniform example that satisfies the boundary conditions. Here we address the question of what other non-uniform solutions exist on this disc.

For now we assume that  $h\neq \pm 2u$ so that $n_3^\star\neq \pm1$, so that $n_3=n_3^\star$ is a circle dividing the target sphere into two regions. Any continuous solution must attain all values on the target space in one of the two regions on the target sphere bounded by $n_3=n_3^\star$. To disambiguate which region, we specify if $n_3$ attains $+1$ or $-1$, meaning it must attain all values in the set $(n_3^\star,1]$ or $[-1,n_3^\star)$ respectively. This motivates the definition:

\begin{definition}
\label{def:moduli-spaces}
The moduli spaces $\mathcal{M}^+_{h,u}$, $\mathcal{M}^-_{h,u}$, are the spaces of all continuous and differentiable solutions to the Euler-Lagrange equations with the standard potential \eqref{eq:standard_pot}, 
\begin{enumerate}[(i)]
    \item on the disc with radius $R_{\rm max}(h,u)$ \eqref{eq:Rn3star} satisfying the boundary conditions $n_3=n_3^*(h,u)$ and $\vec{\nu}$ tangent to the boundary, if $R_{\rm max}(h,u)$ is finite
    \item on the plane, satisfying the boundary conditions at infinity $n_3\to n_3^*(h,u)$, $\vec{\nu}\cdot \hat{\vec{r}}\to 0$, if $R_{\rm max}(h,u)$ is infinite
\end{enumerate} 
attaining $n_3=\pm 1$ respectively in the interior.
\end{definition}
These definitions of the moduli spaces also make sense for the special cases $h=\pm 2u$, but 
$\mathcal{M}^+_{-2u,u}$ is empty, as is $\mathcal{M}^-_{2u,u}$. For $0\leq h \leq 2u$, the moduli spaces formally exist but the solutions they describe are unstable, in the same way as described in Sec.~\ref{sec:static_axisymmetric_solns}. 

Since such solutions are relative homotopies of the axisymmetric solutions, the topological charge of all solutions in $\mathcal{M}_{u,h}^\pm$ is the same as in \eqref{eq:top_charge}.

Note that the boundary condition in case (ii) breaks translation symmetry by picking a preferred origin that lies `at the centre of the path line at infinity', so a fuller moduli space of continuous plane-filling solutions can be found by taking $\mathcal{M}_{h,u}^{\pm}\times \mathbb{R}^2$.

With this definition, we can equate valid static solutions of the Euler-Lagrange equations to a certain function space:
\begin{theorem}\label{Thm_moduli_space}
    For $h\neq \pm 2u$, $\mathcal{M}^+_{h,u}$ and $\mathcal{M}^-_{h,u}$ are in bijection with each other and with the space of $C^1$ functions, $\vec{X}_c$, from: 
    \begin{enumerate}[(i)]
        \item the closed interval $[0,R_{\text{max}}(h,u)]$ to the disc $B(R_{\text{max}})$, with boundary condition $\vec{X}_c(R_{\text{max}})=\vec{0}$, if $R_{\text{max}}(h,u)$ is finite,
        \item the half-line $[0,\infty)$ to the plane, with boundary condition $\lim_{R\to\infty}\vec{X}_c(R)=\vec{0}$, if $R_{\text{max}}(h,u)$ is infinite,
    \end{enumerate}
    whose gradient has a norm always smaller than 1.
\end{theorem}
Informally, the proof follows from the fact that contours of constant $n_3$ of any static continuous solution form a kind of foliation of the domain by circles and a point, which can be parametrized by the positions of the centres of the circles of each radius. We call the function that takes in a radius and outputs the centre of the $n_3$ contour of that radius $\vec{X}_c(R)$, and find that it must belong to the above function space, and conversely that any function belonging to that function space generates a valid solution. 

Formally, first we show that any $\bn$ belonging to $\mathcal{M}_{h,u}^\pm$ is described by a function $\vec{X}_c$ that is in the above function space.

Firstly, the boundary conditions and continuity ensure that $\bn$ attains all values in one of the two spherical caps bounded by $n_3=n_3^\star$. We can consider the preimage of a set of constant $n_3$ in the interval $(-1,n_3^\star)$ or $(n_3^\star,1)$ for $\bn$ in the moduli space $\mathcal{M}^-_{h,u}$, $\mathcal{M}^+_{h,u}$ respectively. It must then be a circle or set of circles inside the domain, of given radius $R(n_3)$ as in \eqref{eq:radius_generic}. So to a given $n_3$ and thus a given $R\in[0,R_{\text{max}})$ (since $R(n_3)$ is invertible), we have in general a set of points $\{\vec{X}_c^1,\vec{X}_c^2\ldots\}$ that are the centres of the these circles, as in the general path line solution \eqref{eq:X_pathline_soln}. 

To form a valid solution, these circles (and points, for $R=0$) must foliate the plane, i.e. not intersect and also fill every point. By definition, these circles fill the plane. Non-intersection of circles means that one of two inequalities is satisfied for two distinct circles:
\begin{align}
    \begin{cases}
        \lvert \vec{X}_c^i(R^A)- \vec{X}_c^j(R^B)\rvert < \lvert R^A-R^B\rvert & \text{circles nested}\\
        \lvert\vec{X}^i_c(R^A)- \vec{X}^j_c(R^B)\rvert> R^A+R^B
& \text{otherwise}    \end{cases}\label{eq:general_non-intersection}
\end{align}

The key claim we must prove is that any point outside a circular path line lies on a circular path line encircling the first path line. Thus, all circles are nested, and to each $R$ we can associate a single $\vec{X}_c(R)$.

To see this, first consider an open neighbourhood $U(\vec{x})$ of a point $\vec{x}$ on a circular path line $C(\vec{x})$ which is not a point,  i.e. $R(n_3(\vec{x}))\neq0$. The curve $C(\vec{x})$ divides $U(\vec{x})$ in two. Let us call the outward normal to the path line $\vec{e}(C(\vec{x}))$. For any point $\vec{x}'$ in $U$ such that $(\vec{x}'-\vec{x})\cdot \vec{e}\geq0$, the corresponding path line $C(\vec{x}')$ must enclose $C(\vec{x})$ (and thus $R(n_3(\vec{x}'))>R(n_3(\vec{x}))$), as the latter inequality in \eqref{eq:general_non-intersection} cannot be satisfied for sufficiently small neighbourhood. 

Let us now consider any $\vec{x}_1$ in the plane outside $C(\vec{x})$. We can define $\vec{x}_0$ as the closest point of $C(\vec{x})$ to $\vec{x}_1$, and define a straight path interpolating between them parametrized by $\sigma$, $\vec{x}(\sigma)=\vec{x}_0 + \sigma (\vec{x}_1-\vec{x}_0)$.

Now we consider the scalar function $\vec{e}(C(\vec{x}(\sigma)))\cdot(\vec{x}_1-\vec{x}_0)$. It must remain positive: to see this, suppose that there existed some $\vec{x}_2$ such that $\vec{e}(C(\vec{x}_2))\cdot(\vec{x}_1-\vec{x}_0)=0$. Considering a point $\vec{x}_3$ between $\vec{x}_0$ and $\vec{x}_2$ in the neighbourhood of $\vec{x}_2$, then $(\vec{x}_3-\vec{x}_2)\cdot\vec{e}=0$ therefore $C(\vec{x}_3)$ encloses $C(\vec{x}_2)$ and thus $\vec{e}(C(\vec{x}_3))\cdot(\vec{x}_1-\vec{x}_0)<0$. That is, the scalar function defined above must be negative in the neighbourhood immediately before a zero, and thus cannot decrease from positive to zero as $\sigma$ is increased.

Therefore, $R$ increases continuously across this line and $C(\vec{x}')$ encloses $C(\vec{x})$. Having established that all contours form one ordered set of nested circles, we know that there is only a single circle of each radius. We can thus define a single-valued map $\vec{X}_c(R)$ that for a given radius $R$ gives the point in the plane that is the centre of the path line with radius $R$, with constant value $n_3$ such that $R=R(n_3)$. For all distinct values of $R$, this function must satisfy the first inequality in \eqref{eq:general_non-intersection}:
\begin{eqnarray}
    \lvert \vec{X}_c(R^A)- \vec{X}_c(R^B)\rvert < \lvert R^A-R^B\rvert\ .
    \label{eq:Xc_condition}
\end{eqnarray}

To proceed it is useful to write the solution $\bn$ corresponding to a given $\vec{X}_c$ in polar-like co-ordinates adapted to $\vec{X}_c$:
\bee
\bn=\begin{pmatrix}
-\sqrt{1-n_3(R)^2}\sin\varphi(\vec{X}_c(R))\\
\sqrt{1-n_3(R)^2}\cos\varphi(\vec{X}_c(R))\\
n_3(R)
\end{pmatrix},
\label{eq:coord_transform_soln}
\eee
where $\varphi(\vec{X}_c)$ means azimuthal co-ordinates with $\vec{X}_c$ as the centre, and $R^2=(x_1-X_{c,1}(R))^2 + (x_2-X_{c,2}(R))^2 $. This equation has $R$ on both sides and is thus hard to effect as a co-ordinate transformation for a general function. 
The constant function $\vec{X}_c(R)=\vec{0}$ gives the axisymmetric solutions in Sec.~\ref{sec:static_axisymmetric_solns}, while in general a solution will have circles of constant $n_3$ at different displacements within.

Since $\vec{X}_c(R)$ is Lipschitz, it is differentiable except at isolated points. If we assume that $\vec{X}_c(R)$ does not have a derivative at a point, we find that $\bn$ is not differentiable. To see this, we use the fact that locally, such a point of non-differentiability can be represented by
\begin{eqnarray}
    \vec{X}_c(R) = \begin{cases}
        v_A R \vec{e}_A & 0\leq R\leq R_\star\\
        v_A R_\star  + v_B(R-R_\star)\vec{e}_B & R_\star \leq R \leq R_{\text{max}}
    \end{cases},
\end{eqnarray}
with $0<R^\star< R_{\text{max}}$, $\vec{e}_{A,B}$ arbitrary unit vectors in the plane and $0\leq v_{A,B}<1$. This form is simple enough to find the co-ordinate transform explicitly, and thus compute $\nabla\bn$ and find it is discontinuous unless $v_A=v_B$ and $\vec{e}_A=\vec{e}_B$, i.e. if there is no discontinuity in $\vec{X}_c'(R)$.
Since $\vec{X}_c(R)$ must be differentiable, equation \eqref{eq:Xc_condition} becomes
\begin{eqnarray}
    \lvert \vec{X}_c'(R)\rvert<1\ .
\end{eqnarray}
In case (i), we have the boundary condition on the map $R_{\text{max}}$ to $\vec{0}$, the origin of the disc. In case (ii), the boundary condition tells us that $\lim_{R\to\infty}\vec{X}_c=\vec{0}$. The constraints on $\bn$ thus give us the constraints on $\vec{X}_c$ described in Thm.~\ref{Thm_moduli_space}.

Conversely, we can show that any function $\vec{X}_c$ in this space gives a valid continuous and differentiable static solution: we can think of the field configuration corresponding to a given $\vec{X}_c$, $\bn(\vec{x};\vec{X}_c)$ being a composition $\bn(\tilde{X}_c(\vec{x}))$ where $\tilde{X}_c:\mathbb{R}^2\to\mathbb{R}^2$ maps the circle of points with centre $\vec{X}_c$, radius $R(\vec{X}_c)$ to the circle of the same radius with centre 0 by translation. Since the circles of each $R$ do not intersect, $\tilde{X}_c$ is well-defined. 
Moreover we can go to polar co-ordinates after the transformation $\tilde{X}_c$, $(\tilde{r},\tilde{\phi})$, and explicitly show that the derivative with respect to $\vec{x}$ of $\tilde{r}$ and $\tilde{\phi}$ (away from $\tilde{r}=0$) is defined when $\vec{X}_c'(R)$ is defined. It follows that $\tilde{X}_c$ is differentiable, and thus $\bn$ is differentiable. The function $\tilde{\phi}(\vec{x})$ can be checked to be differentiable except at $\tilde{r}=0$ by moving to complex co-ordinates. To show $\tilde{r}(\vec{x})$ is differentiable we differentiate the implicit equation $\tilde{r}^2=(x_1-X_{c,1}(\tilde{r}))^2 + (x_2-X_{c,2}(\tilde{r}))^2 $
and solve for the derivative:
\begin{eqnarray}
    \partial_{\vec{x}}\tilde{r} = \frac{(\vec{x}-\vec{X}_c)\cdot\partial_{\vec{x}}\vec{x}}{\tilde{r}+(\vec{x}- \vec{X}_c)\cdot\vec{X}'(R)}\ .
\end{eqnarray}
The denominator of this fraction cannot be zero, since on the one hand $\tilde{r} = \lvert\vec{x}-\vec{X}_c\rvert $, and on the other $\lvert(\vec{x}-\vec{X}_c)\cdot\vec{X}_c'(R)\rvert\leq\lvert\vec{x}-\vec{X}_c\rvert \lvert \vec{X}_c'(R)\rvert<\lvert\vec{x}-\vec{X}_c\rvert$. Therefore $\tilde{r}(\vec{x})$ is differentiable. We thus establish a one-to-one mapping between solutions to the Euler-Lagrange equations and functions in the function space described in Thm.~\ref{Thm_moduli_space}.
This concludes the proof.

\subsection{Effective string-like model}
\begin{figure}[H]
    \centering
    \includegraphics[width=0.85\linewidth]{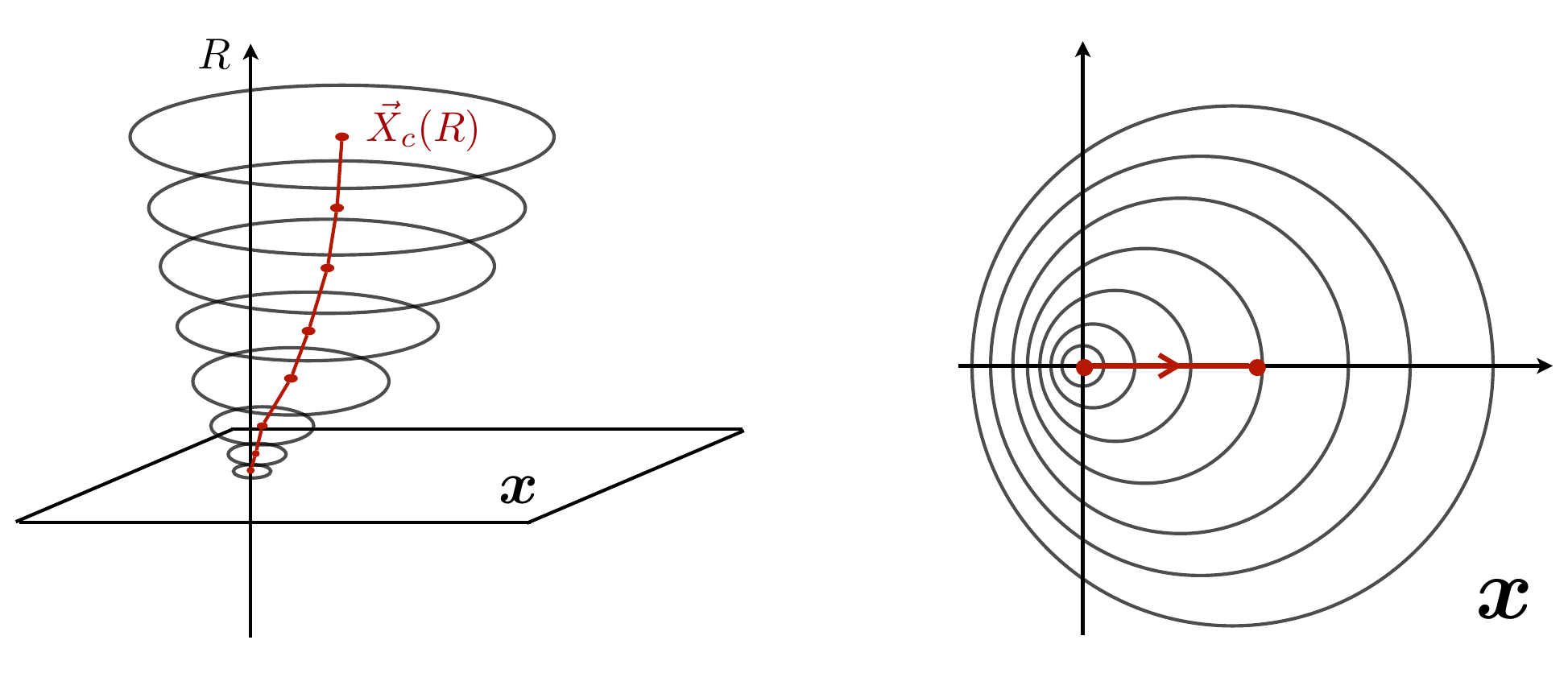}
    \caption{Pictorial illustration of the string-like model.}
    \label{fig:string}
\end{figure}

As we noticed above in Eq.~\eqref{eq:Xc_condition}, the moduli space can be parametrized by  $\vec{X}_c(R)$, the line of the centres depending on $R(n_3)$ as in Eq.~\eqref{R_n3}. From this perspective, the moduli space is a sort of ``string model'' with $R$ parametrizing the string and $\vec{X}_c$ giving its embedding in space. This is shown pictorially in Fig.~\ref{fig:string}. In the chiral magnet model without the Heisenberg term, all the  configurations \eqref{eq:Xc_condition} have the same energy; thus the effective string would have an infinite square-well elastic potential, tension-less up to some maximum extension. At first level of approximation, we can compute the effect of the Heisenberg contribution by evaluating $E_2$ on the solutions \eqref{eq:coord_transform_soln} of the model without Heisenberg interaction. We know that only the $\SO(2)$ invariant solutions with $\vec{X}'_c(R)=\vec{0}$ will be the true minima of the energy \eqref{eq:energy}.

The $E_2$ term reads
\begin{align}
	E_2[\Theta, \Psi] = \int \d^2x \, \left(
	\frac{1}{2}(\partial_i\Theta)^2
	+ \frac12\sin^2\Theta (\partial_i \Psi)^2 \right),
 \end{align} 
with $\bn=(\sin\Theta\cos\Psi,\sin\Theta\sin\Psi,\cos\Theta)$
and the fundamental state is of the form: $\Theta = \Theta(r)$ and $\Psi = \phi + \delta$, where $\delta=\frac\pi2$ for the Bloch-type DMI.
Note that for the  supercompacton~\cite{Bolognesi:2024mjs}, $E_2$ has a divergence near the boundary where $\lim_{r \to R_{\rm max}}\Theta'(r) = \infty$. Now we perturb the solution with a modulation of the centres $\vec{X}_c(R)$ and evaluate the variation $\Delta E_2$. We use polar co-ordinates centred in each circle of radius $r$. At first order the distance between the two circles is $\delta r(\phi) = \big(1+ |\vec{X}'_c(R)| \cos(\phi-\phi_{\vec{X}_c})\big)\d r$
where $\phi_{\vec{X}_c}= \arctan ({X}'_{c,2}/{X}'_{c,1})$. The area is in fact preserved as $\int \d \phi\delta r(\phi) = 2\pi \d r$. The gradient $\partial_i \Psi$ is not affected by the circle's displacement, thus $\Delta E_2$ gets a contribution only from the term with $\partial_i\Theta$. The result is then
\bee
\Delta E_2 = \int \d r \d \phi \;r \frac{1}{2}(\partial_r\Theta)^2 \left(\frac{1}{1+ |\vec{X}'_c(R)| \cos(\phi-\phi_{\vec{X}_c})}-1\right).
\eee
(There is one factor $\delta r$ from the measure, and one $\frac{1}{\delta r ^2}$ from the gradient.)
For small displacements, 
\bee
\Delta E_2 = \int \d r\; r \frac{1}{8}(\partial_r\Theta)^2  |\vec{X}'_c(R)|^2 \ .
\eee
This is essentially a tension term for the string, where the tension density is $r \frac{1}{8}(\partial_r\Theta)^2 $ which is dependent on $r$. If $R \to R_{\rm max}$ sufficiently fast, such that $\vec{X}'_c(R) \to 0$, then $\Delta E_2$ can be finite, even for a supercompacton where $E_2$ diverges.

\subsection{Explicit solutions}

\begin{figure}[!htp]
\centering
\begin{tabular}{p{0.25\linewidth}p{0.25\linewidth}p{0.25\linewidth}}
linear & quadratic & trigonometric
\end{tabular}
\mbox{
\includegraphics[width=0.25\linewidth]{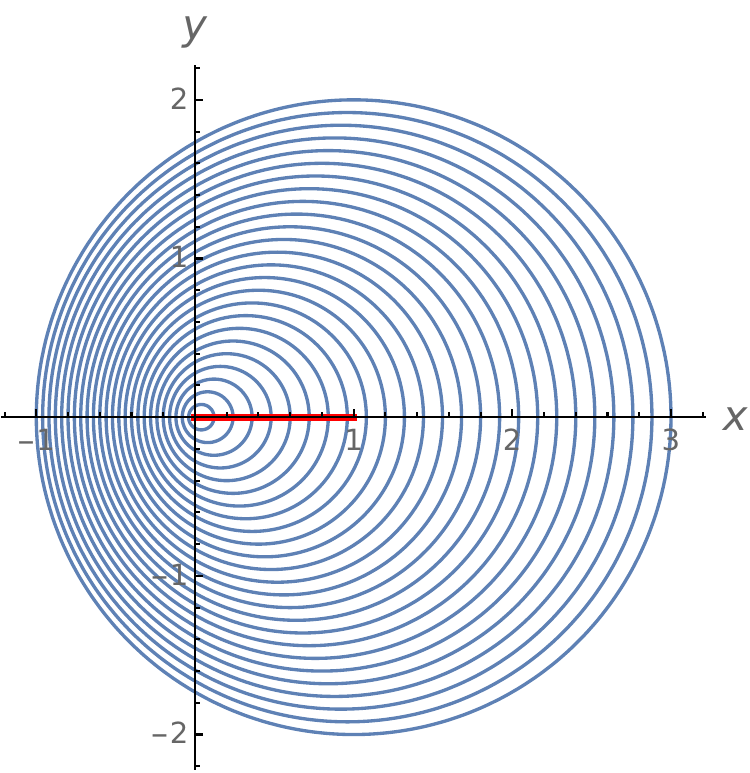}
\includegraphics[width=0.25\linewidth]{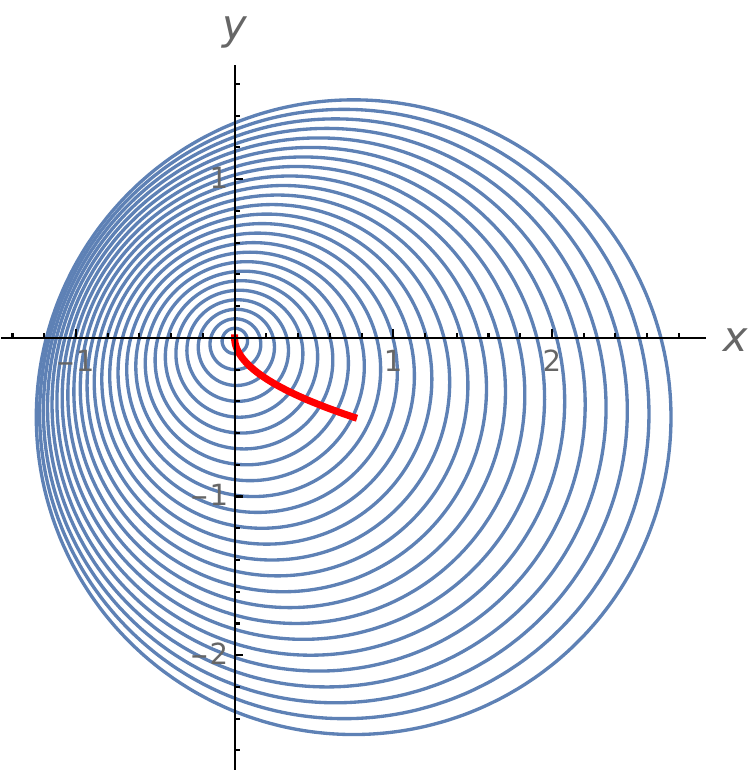}
\includegraphics[width=0.25\linewidth]{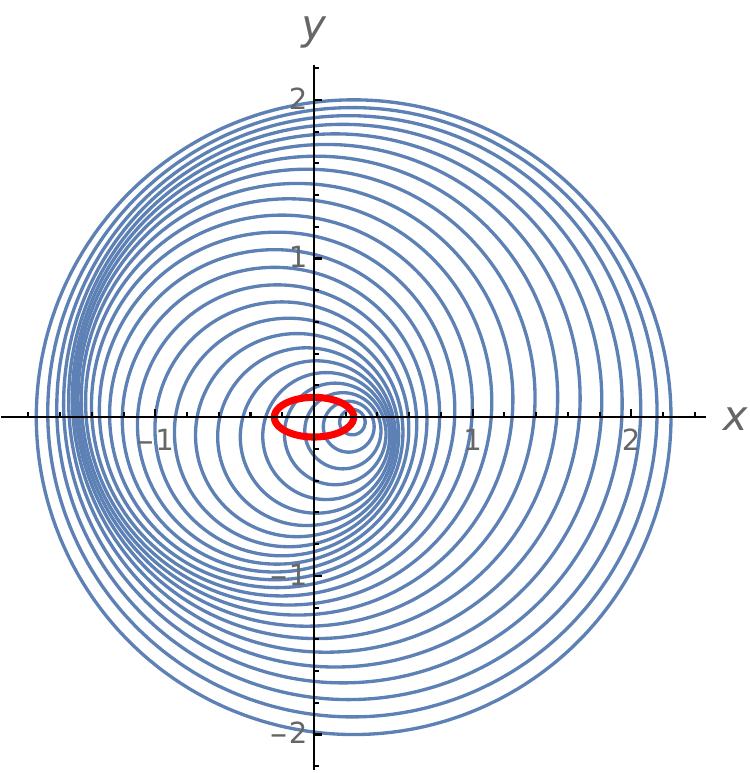}}\\
\mbox{
\includegraphics[width=0.25\linewidth]{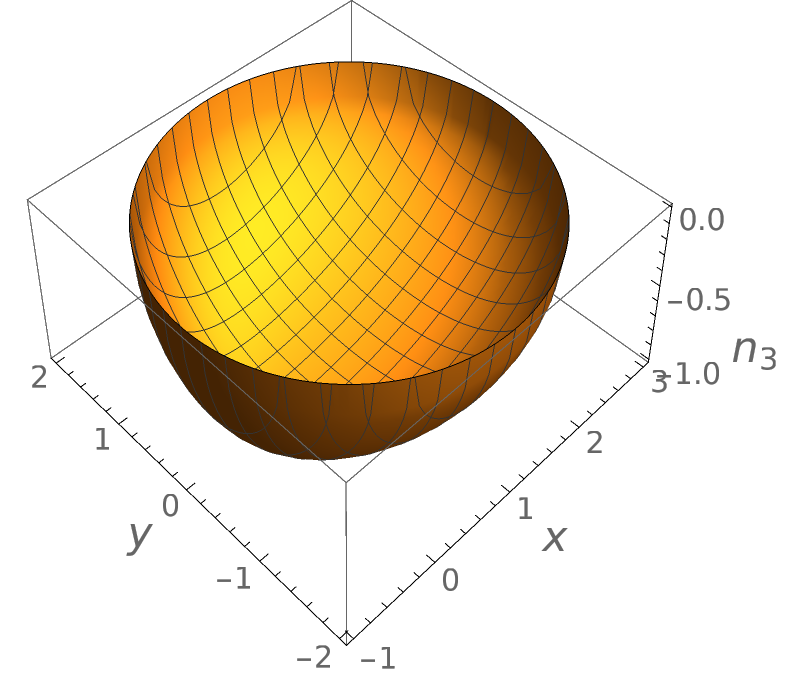}
\includegraphics[width=0.25\linewidth]{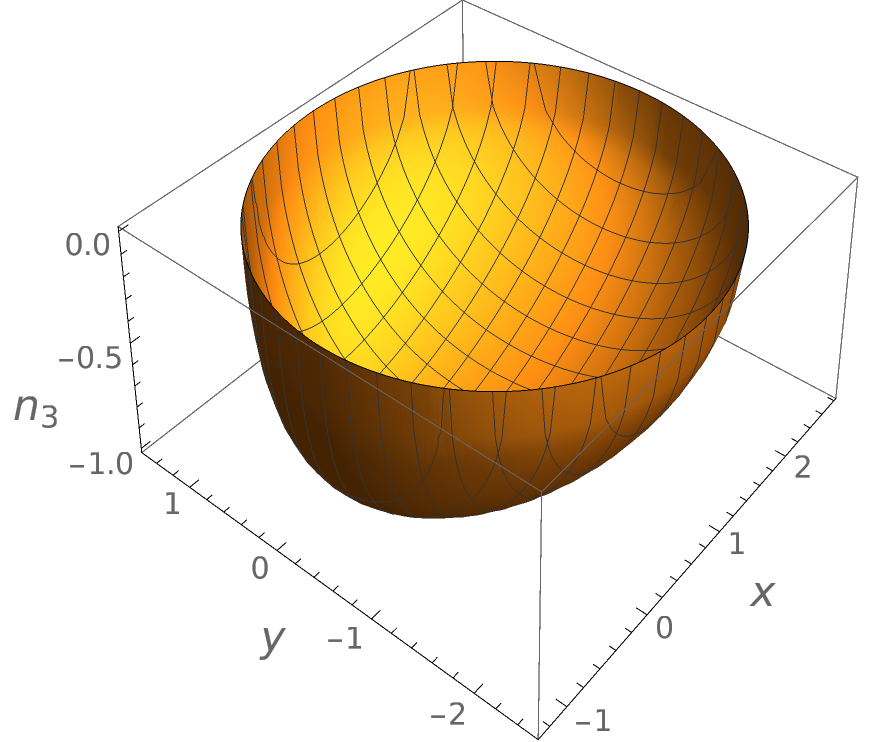}
\includegraphics[width=0.25\linewidth]{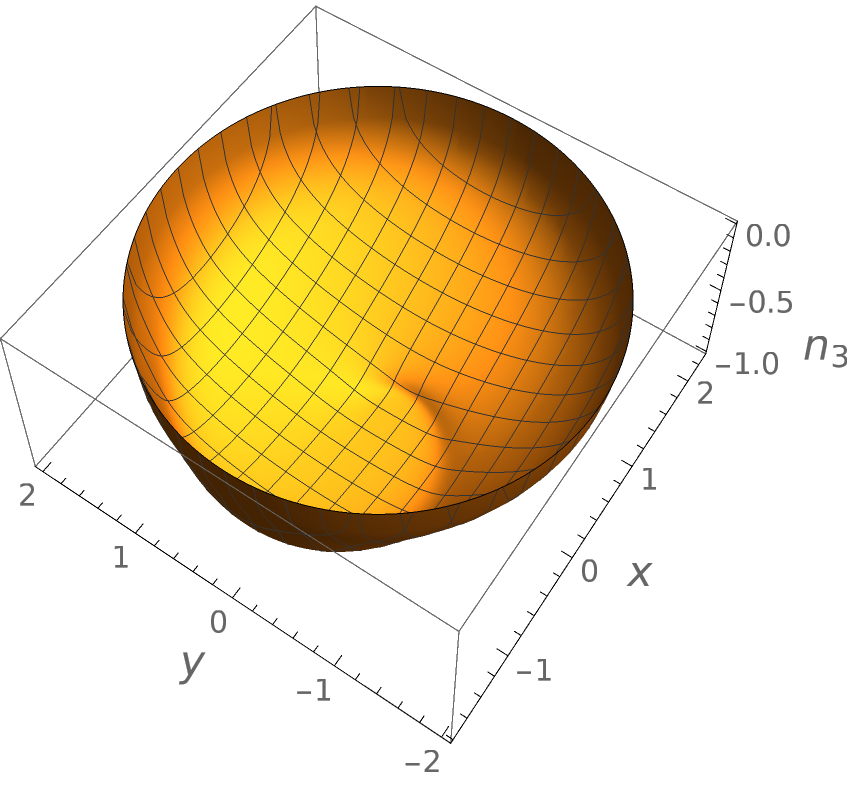}}\\
\mbox{
\includegraphics[width=0.25\linewidth]{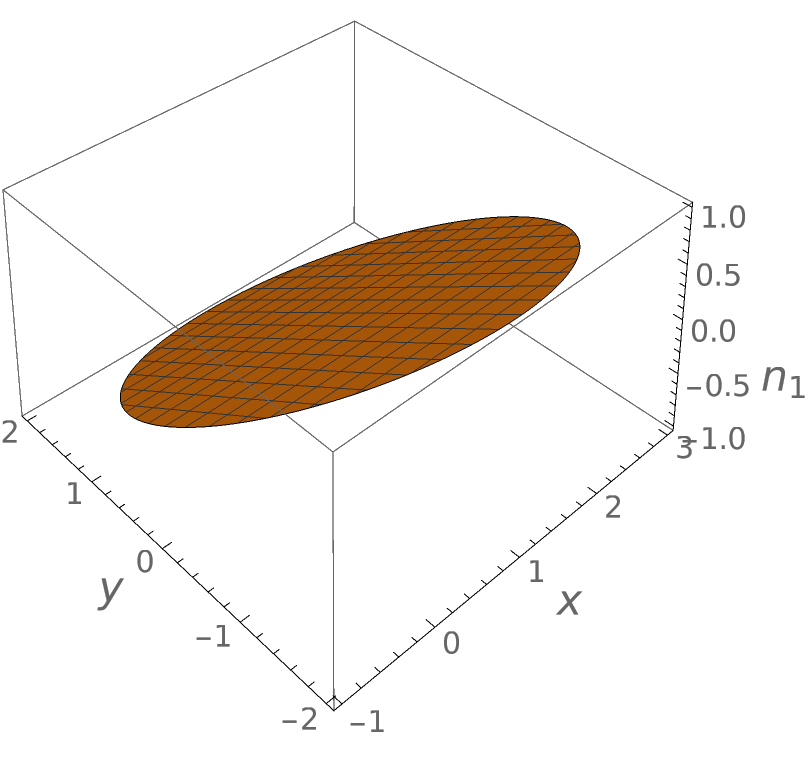}
\includegraphics[width=0.25\linewidth]{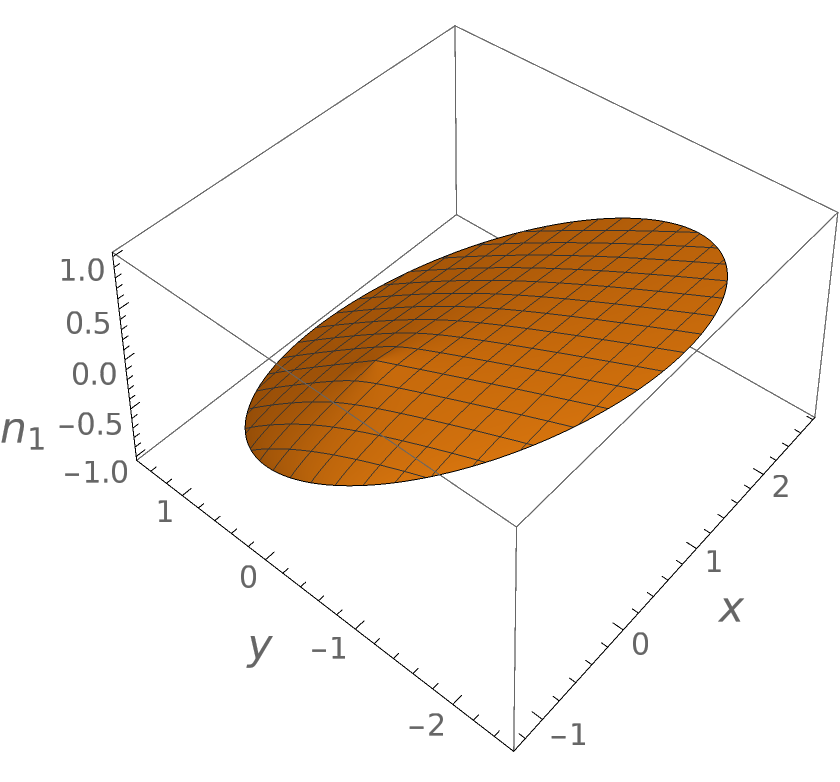}
\includegraphics[width=0.25\linewidth]{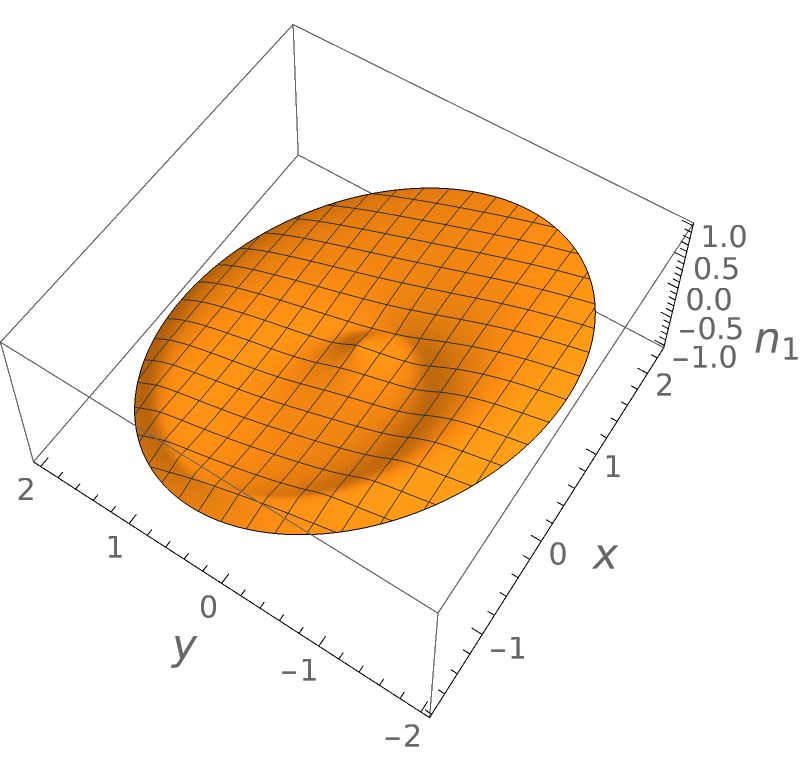}}\\
\mbox{
\includegraphics[width=0.25\linewidth]{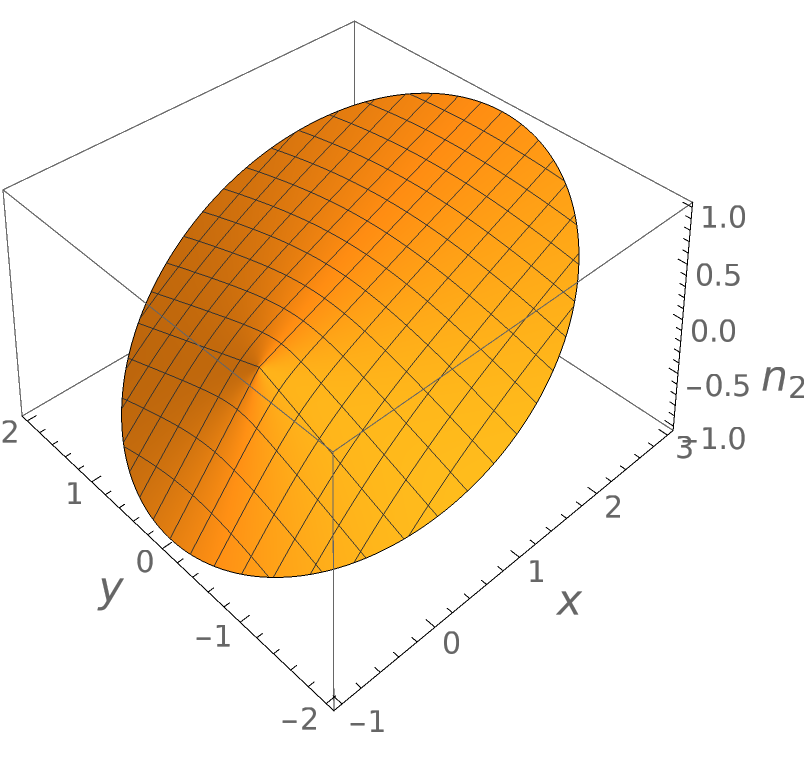}
\includegraphics[width=0.25\linewidth]{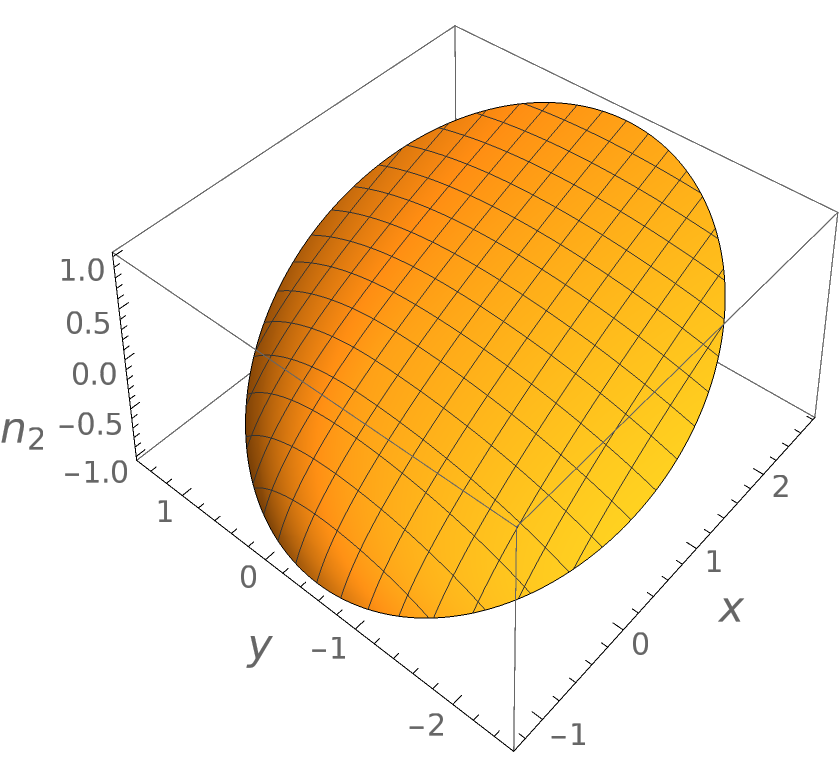}
\includegraphics[width=0.25\linewidth]{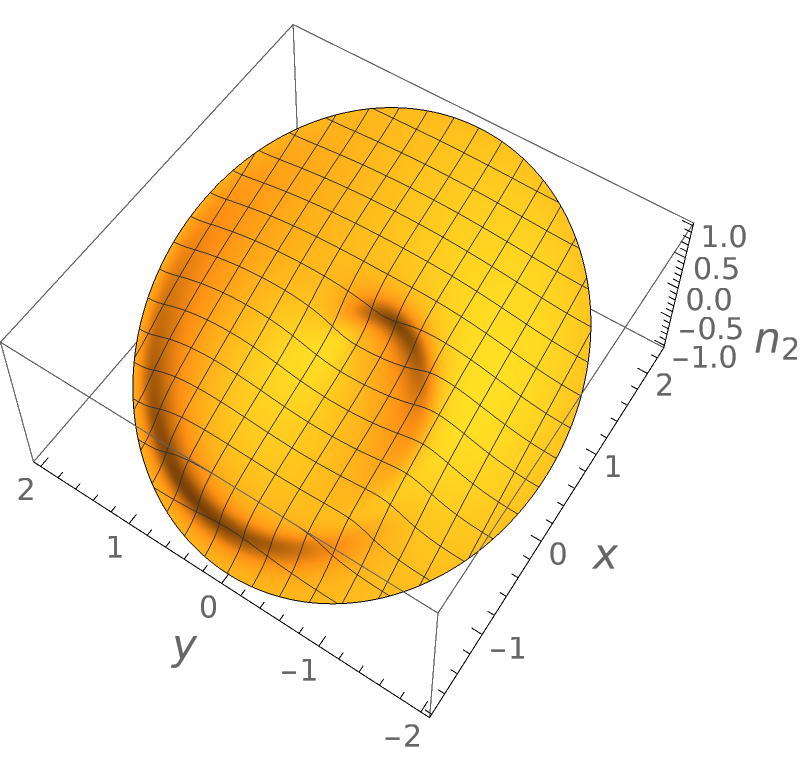}}
\caption{Three solutions of restricted magnetic Skyrmions, with different non-trivial ``string motion'' in the case of the Zeeman potential $h=1$, $u=0$, giving rise to a supercompacton. The four rows correspond to the collection of circles and the fields $n_3$, $n_1$ and $n_2$, respectively. In this figure $k=1$. }
\label{fig:solns}
\end{figure}

We will now illustrate a few examples of explicit solutions.
The solutions are shown in Fig.~\ref{fig:solns}.
The three solutions shown in the figure correspond to the following string motions:
\begin{align}
\textrm{linear}: X_c(R)&=-\frac{R}{2},\\
\textrm{quadratic}: X_c(R)&=\frac{-3R^2}{16}+\frac{\i R}{4},\\
\textrm{trigonometric}: X_c(R)&=-\frac14\cos(\pi R)+\frac{\i}{8}\sin(\pi R),
\end{align}
where we have chosen $h=1$, $u=0$ leading to supercompactons with $R\in[0,2]$.
We checked that the evaluation of the equation of motion \eqref{fluid_flow_eq} vanishes to numerical precision of the derivatives computed numerically on an interpolated field in the \texttt{Mathematica} package.

\section{Dynamics of breather-like solutions}
\label{sec:dynamics}

We can also understand dynamical solutions through these characteristics, by relaxing the condition that flows are stationary. Note that from \eqref{eq:X_pathline_soln} we can define $\omega(n_3)=-U'(n_3)$, which represents the frequency with which a single spin, followed  along its trajectory, returns to the initial point. 

Let us then consider the set of points where $n_3$ is equal to some constant. Suppose that initially this is a circle of radius $R_0$, with $\vec{\nu}$ tangent to the circle at all points, and that $R_0>R(n_3)$. As time evolves, the curve along which $n_3$ is constant will contract, reaching a minimum radius $R_{\text{min}} = 2R(n_3)-R_0$ and then expand, reaching the radius $R_0$ again at time $\frac{2\pi}{\omega(n_3)}$. If we can combine characteristics for each value of $n_3$, then we will produce a breathing compacton.

Systematically, we should consider a set of characteristics with equal $n_3$, parametrized by periodic variable $\bar{\sigma}=\frac{\sigma}{2\pi R(n_3)}$, starting tangent to a circle of radius $R(n_3)+\Delta(n_3)$ at time $\Phi(n_3)/\omega(n_3)$ (that is, we allow different oscillations to have different phases $\Phi$). We can take $\Delta>0$ without loss of generality. Note that $\Delta$ is the initial perturbation profile, it is not necessarily small. As before, we have complex co-ordinates $X$ giving the position of the characteristic and $\nu$ giving the tangent direction and speed of the characteristic with respect to time. Since we are now considering a one-parameter family of characteristics, $X$ and $\nu$ are functions of two variables, $t$ and $\bar{\sigma}$:
\bee
X(t,\bar{\sigma}) = R(n_3) e^{\i\omega(n_3)t + \i\bar{\sigma}} + \Delta(n_3) e^{\i\bar{\sigma}+\i\Phi(n_3)} \ ,\quad \quad \nu(t,\bar{\sigma}) = \i e^{\i\omega(n_3)t + \i\bar{\sigma}} \ .
\label{eq:breathing_characteristics}
\eee
We see immediately that at fixed $t$, the set of points with $n_3$ equal is a circle centred at 
$X_c=0$ with radius
\bee
\mathcal{R}(n_3,t) = \sqrt{R(n_3)^2 + \Delta(n_3)^2 + 2R(n_3)\Delta(n_3)\cos(\omega(n_3)t-\Phi(n_3))} \ .
\eee

To avoid this $\mathcal{R}$-circle to not collapse to zero size, we require $\Delta<\lvert R\rvert$. 

To construct an extended solution, we want to put together the circles of different constant $n_3$ inside each other. In general, this is not possible since the period of oscillation $\frac{2\pi}{\omega(n_3)}$ will be different for each $n_3$, and thus we cannot avoid the different circles colliding, at which point our field becomes discontinuous. However, with pure Zeeman interaction, that is the potential \eqref{eq:standard_pot} with $\anis=0$, $\omega(n_3)=\zee$ and thus all circles have the same period of oscillation. Therefore, to look for breathing solutions, we must specialize to a model with only Zeeman interaction in the potential.

Even if all circles oscillate in radius at the same time, they have different amplitudes that depend on their initial radius, so we must choose those initial radii so that the circles never collide, or shrink to zero radius.

This means finding $\Delta(n_3)$, $\Phi(n_3)$ such that
\bee
\begin{cases}
	\Delta(-1)=0 \\
	 \Delta<R\\
	\partial_{n_3} \mathcal{R}^2 >0 \ , \quad \quad 0<t<\frac{2\pi}{\zee}\ .
\end{cases}
\label{eq:Delta_constraints}
\eee
The third equation simplifies to
\begin{equation}
	\Delta \Delta'(n_3)+ R R'(n_3) +(R\Delta'(n_3) + \Delta R'(n_3))\cos(ht-\Phi)  -R\Delta\Phi'(n_3)\sin(ht-\Phi)>0\ .
\end{equation}
We should thus find the strongest constraint for all $t$, which will be
\begin{equation}
\Delta \Delta'(n_3)+ R R'(n_3)  -\sqrt{(R\Delta'(n_3) + R'(n_3)\Delta)^2+ (R\Delta\Phi'(n_3))^2}>0\ .
\end{equation}

\begin{figure}
	\begin{center}
		\begin{tikzpicture}[scale=0.8]
			\draw (0,0) circle (5);
			\draw[dashed] (1,0) circle (4) ;
			\draw (0,0) circle (3);
			\draw[<->,rosso] (1,0) --node[anchor=south] {\small $R(n_3)$} (5,0) ;
			\draw[<->,rosso] (0,0) --node[anchor=east] {\small $R(n_3)+\Delta(n_3)$} (-3,-4) ;
			\draw[<->,rosso] (0,0) --node[anchor=west] {\small $R(n_3)-\Delta(n_3)$} (-2,2.236) ;
			
			\draw[->,thick]        (0,-5)   -- (0.5,-5);
			\draw[->,thick]        (0,5) node[anchor=south] {\small $n_3$ constant at $t=\frac{\Phi(n_3)}{h}$}   -- (-0.5,5);
			\draw[->,thick]        (5,0)   -- (5,0.5);
			\draw[->,thick]        (-5,0)   -- (-5,-0.5);
			
			\draw[->,thick]        (1,4) node[anchor=south] {\small example path line of individual spin}   -- (0.5,4);
			\draw[->,thick]        (1,-4)  -- (1.5,-4);
			\draw[->,thick]        (-3,0)   -- (-3,-0.5);
			\draw[->,thick]        (0,3) node[anchor=south] {\small $n_3$ constant at $t=\frac{\Phi(n_3)+\pi}{h}$}   -- (-0.5,3);
			\draw[->,thick]        (0,-3)   -- (0.5,-3);
					\end{tikzpicture}
		\caption{\small Constant $n_3$ contours at different points in time. Note that while the magnetization is always tangent to its path line, for $ht-\Phi\neq n\pi$ it is not tangent to the circle of constant $n_3$, in contrast to the axisymmetric static solution.}
		\label{fig:illustration2}
	\end{center}
\end{figure}
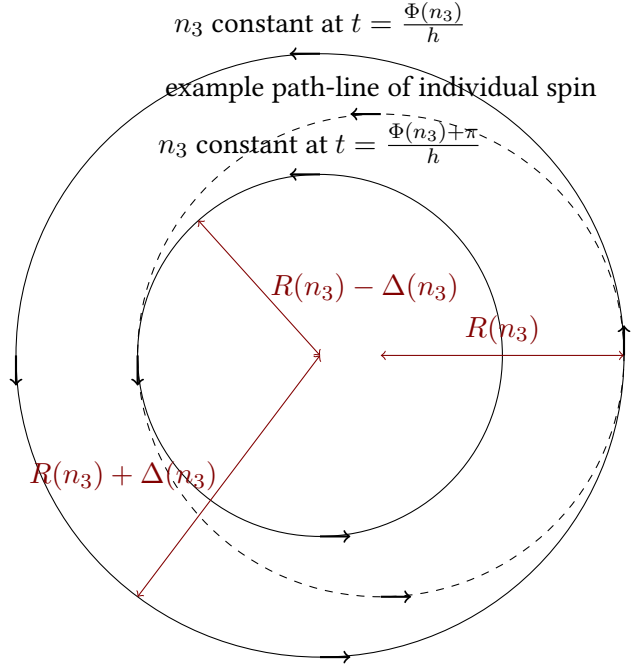

This constrains $\Delta(R)$, $\Phi(R)$ in a similar way to the way that $\vec{X}_c(R)$ is constrained in \eqref{eq:Xc_condition}. 
This constraint can be written as
\begin{equation}
\lvert\Delta'(R)\rvert<1,\,\lvert\Phi'(R)\rvert<\sqrt{(1-\Delta'(R)^2)(\lvert\Delta\rvert^{-2}-\lvert R\rvert^{-2})}\ .
\end{equation}
For an illustration, see Fig.~\ref{fig:illustration2}.

Note that it is valid, within these constraints, to have a $\Delta$ that is supported only in a subset of $[-1,n_3^\star]$ or $[n_3^\star,1]$. Physically, this corresponds to a Skyrmion that only ``breathes'' within a certain annulus, and it is otherwise static. 

In contrast to the static case, where attainment of the inequality meant an invalid solution, here $\Delta =\lvert R\rvert$ or $\lvert \Delta'\rvert =\lvert R'\rvert$ at some $n_3$ will in general give the development of a singularity in finite time. This could occur at a point, at the centre of the Skyrmion, or along a circle. 

These solutions might prove useful as starting points for understanding periodic solutions and finite-time collapse in the full model with symmetric exchange interaction.

Note that the $\U(1)$ action where we rotate both $\bn$ and $\vec{x}$ in opposite directions is a symmetry of the energy, and this symmetry is preserved by Landau-Lifshitz dynamics. Thus we can consistently constrain the Landau-Lifshitz equation to the most general $\U(1)$ ansatz:
\bee
\bn=\begin{pmatrix}
	\sin\Theta(r)\cos(\phi+\delta(r))\\
	\sin\Theta(r)\sin(\phi+\delta(r))\\
	\cos\Theta(r)
\end{pmatrix} \ ,
\label{eq:axisymm-ansatz}
\eee
where $(r,\phi)$ are polar co-ordinates in the plane. Solving the resulting coupled ODEs for $\Theta(r)$ and $\delta(r)$ should then give the same set of solutions.

We can get $\delta$ from \eqref{eq:breathing_characteristics}, more conveniently expressed as a function of $n_3$ (we find $n_3(r,t)$ by inverting $\mathcal{R}(n_3,t)$):
\bee
\delta(n_3(r,t),t)=\frac{\pi}{2}- \arg\left(1 + \frac{\Delta(n_3(r,t))}{R(n_3(r,t))}e^{\i(\Phi(n_3(r,t))-ht)}\right).
\eee
If $\delta(r)$ is constant, it is the `Skyrmion helicity' variable commonly referred to in the literature. This equation reflects the coupling between helicity and breathing that has been observed in collective co-ordinate approximations for breathing magnetic Skyrmions~\cite{mckeever2019breathing}, where an ansatz is assumed with $\delta$ independent of $r$. We can find special solutions with such $\delta$ when $\Delta/R$ and $\Phi$ are constants independent of $n_3$.

However yet again, we can construct non-axisymmetric breathing solutions that we would not find by this method, now allowing our families of characteristics to have a centre that continually changes with $n_3$:
\begin{eqnarray}
&&X(t,\bar{\sigma}) = R(n_3) e^{\i\omega(n_3)t + \i\bar{\sigma}} + \Delta(n_3) e^{\i\bar{\sigma} + \i\Phi(n_3)} + X_c(n_3) \ , \\
&&\nu(t,\bar{\sigma}) = \i e^{\i\omega(n_3)t + \i\bar{\sigma}+\i\Phi(n_3)} \ .
\end{eqnarray}
This then gives a constraint connecting $\Delta(n_3)$, $\Phi(n_3)$, $X_c(n_3)$ and the given function $R(n_3) $ that unifies the axisymmetric breather (where $X_c'=0$) and the non-axisymmetric static solution (where $\Delta=0$). To derive this we should modify our constraint to
\bee
\begin{cases}
	\Delta(-1)=0 \\
	\lvert \Delta\rvert<R\\
	\partial_{n_3} \mathcal{R} >\lvert X_c'(n_3)\rvert \ , \quad \quad 0<t<\frac{2\pi}{\zee}\ .
\end{cases}
\eee
The third constraint can be rewritten as
\begin{eqnarray}
&&RR'(n_3) + \Delta\Delta'(n_3) + (R'(n_3)\Delta+R\Delta'(n_3))\cos(ht-\Phi) + R\Delta\Phi'(n_3)\sin(ht-\Phi)\\ &&\mathop-\sqrt{R^2+\Delta^2+2R\Delta\cos(ht-\Phi)}\lvert X_c'(n_3)\rvert >0\ . \nonumber
\end{eqnarray}
However in contrast to the previous case, eliminating $t$ by finding the strongest constraint involves solving an equation of the form $\sin(x)+c_1\sin(2x+c_2)=0$, which is unwieldy to write in closed form.

\section{Static solutions in 3D}
\label{sec:3d}

We can generalize to a 3D chiral magnet energy which is essentially the same
\bee
E(\bn) = \int \d^3x \Big( k\bn\cdot(\nabla\times\bn) + V(\bn) \Big)\ ,
\eee
the key difference being that the DMI now includes derivatives with respect to $x_3$, and thus from the fluid point of view, spins are advected by the vector field $\bn$ itself, rather than its projection into the plane.

The rest of the argument is very similar, but now the trajectory $\boldsymbol{X}(t)=(X_1,X_2,X_3)$ as we follow a single spin is a helix, with $(X_1,X_2)$ as before and 
\begin{equation}
X_3(t)=\frac{n_3}{2k} t\ .
\end{equation}
This therefore means that a static solution in 3D must be assembled from these helices where $n_3$ is constant, with $n_3=\pm1$ the degenerate case where the helix is a vertical line. These helices have radii given by $R(n_3)$ as before and length scale in the $x_3$ direction,
\begin{equation}
L(n_3)= \frac{2k}{n_3}\ .
\end{equation}
In the previous case, the circles of constant $n_3$ had two degrees of freedom, in terms of the position of the centre. These helices have two translational degrees of freedom, two orientational degrees of freedom and one periodic degree of freedom corresponding to translation along the helix.

We can certainly arrange solutions whose 2D cross sections are identical to the static solutions above, with trivial dependence in the $x_3$ direction, i.e.~``compacton tubes''. It may also be possible to arrange compacton tubes with some $x_3$ modulation.

However, it will not be possible to assemble a static Hopfion solution: a Hopfion by definition exists within a uniform background $\bn=\bN$ some minimum of $V$, and its topology is defined with respect to that. The Hopfion attains all values of $\bn$ in the interior of its domain. In particular, it attains a value of $\bn$ which is a local maximum of $V$. If a Hopfion existed in this model, if it attained this $\bn$ at any point it would have to attain it along a straight line through that point, which would intersect with the uniform background, giving a contradiction. A Hopfion may exist dynamically, however.

\section{Conclusion}
\label{sec:conclusion}

In this paper we consider the limit of the magnetic Skyrme theory -- the theory of magnetic textures in chiral magnetic materials, where the Heisenberg exchange interaction is absent or negligible. 
Formulating the equations in terms of a fluid-flow equation, we find the interesting result that an enlarged moduli space emerges.
The fluid lines are all circles, but due to the missing Heisenberg energy term, the circles can be moved at will -- as long as they do not intersect one another. 
This leads to the interpretation that the moduli space is that of a ``string'' with the speed limit, corresponding to the circles being non-intersecting.
We give a few explicit examples of exact solutions that are constructed in this way and confirm that they obey the equations of motion for the fields.
We further go on and contemplate the case of time-dependent solutions that correspond to breathers. 
Finally, we conclude that no knotted solitons or Hopfions exist in this model.

As a future direction, one could investigate whether a moduli space exists in the case of the higher-dimensional generalization of the magnetic Skyrmion, recently put forward in Ref.~\cite{Gudnason:2024opf}.

It would also be interesting to study other first-order solitonic systems using the assembly of fluid lines, as used in this paper.

In the discussion of breather-like solutions, we found time-periodic 
solutions and solutions that develop a singularity in finite time. With the introduction of an infinitesimal Heisenberg energy term, these could be used to understand periodic solutions and singularity formation in physically realistic chiral magnets.

All conclusions in this paper followed from looking for continuous solutions to the equations of motion, and thus often found configurations only on a finite disc. More generally, if we allow discontinuous weak solutions, we would expect to find solutions with the whole plane as a domain, but we must deal with the behaviour of the boundary where $\bn$ is discontinuous, using the theory of shock and rarefaction waves. Since the shock condition is trivially satisfied if the fluid flow on both sides of the shock are parallel to the shock boundary, we can look for continuous solutions on subsets of the plane with boundary conditions that $\vec{\nu}$ is parallel to the boundary. These can be assembled to create discontinuous solutions in the whole plane, but conversely we cannot expect to find all discontinuous solutions in the plane by assembling these particular `self-contained' continuous solutions.

\subsection*{Acknowledgements}

The work of B.~B.-S. is supported by the ERC starting grant SINGinGR (Grant No.~101078061), under the European
Union's Horizon Europe program for research and innovation. The work of S.~B. is supported by the INFN special research project
grant ``GAST'' (Gauge and String Theories).
S.~B.~G. thanks the Outstanding Talent Program of Henan University and
the Ministry of Education of Henan Province for partial support.
The work of S.~B.~G.~is supported by the National Natural Science
Foundation of China (Grant No.~12071111) and by the Ministry of
Science and Technology of China (Grant No.~G2022026021L).

\appendix
\renewcommand{\theequation}{A\arabic{equation}}
\setcounter{equation}{0}

\section{The restricted model in complex co-ordinates}

Let us consider the static energy density \eqref{eq:energy}. We now introduce complex disk co-ordinates, such that $\nu=n_1+\i n_2$, $\nu\nub\leq1$
and $n_3=\pm\sqrt{1-\nu\nub}$, together with $\pm$ covering the north and south
part of the 2-sphere.
The derivative of the standard potential \eqref{eq:standard_pot} is
\beq
-U'(n_3) = \zee + 2\anis n_3\ .
\eeq
Inserting the complex co-ordinates and changing the co-ordinates of the
plane to $z=x^1+\i x^2$ with its complex conjugate, we arrive at $E=\int\d^2x\;\calE$:
\beq
\calE =
\pm\i k\frac{2\pb\nub - 2\p\nu + \nub^2\pb\nu - \nu^2\p\nub + \nu\nub(-\pb\nub + \p\nu)}{2\sqrt{1-\nu\nub}}
+ U(\pm\sqrt{1-\nu\nub})\ ,
\eeq
with the lower sign corresponding to $n_3\leq0$ and the upper sign to $n_3\geq0$.
For the supercompacton, the lower sign is sufficient as the other patch of the 2-sphere does not exist.
The energy is real, so it suffices to write the equation of motion for $\nub$:
\beq
\left(2\i\nu\p + 2\i\nub\pb - \frac{U'}{k}\right)\nu = 0\ ,
\eeq
which is indeed very elegant and independent of the sign $\pm$ choosing the patch. 
A simple solution that covers the north patch, suitable for the Zeeman potential which has $U'=-h$, is given by
\beq
\nu = \nu(z) = \frac{h}{2k\i}z\ .
\eeq
Including $u\neq0$ introduces a square root into the equation, making it more cumbersome.

\bibliographystyle{JHEP}
\bibliography{biblio}
\end{document}